\documentclass[twocolumn,aps,prl,showpacs,floatfix]{revtex4}
\usepackage{graphicx}
\usepackage{times}
\usepackage{nicefrac}
\usepackage{amsmath}
\usepackage{amsfonts}
\usepackage{amssymb}
\usepackage{amsthm}
\usepackage{epsf}
\usepackage{bm}
\usepackage{times}

\usepackage{dcolumn}

\newcolumntype{.}{D{x}{}{-1}}

\newcommand{\bfr}{{\bf r}}

\newcommand{\bfp}{{\bf p}}
\newcommand{\bfA}{{\bf A}}
\newcommand{\bfD}{{\bf D}}
\newcommand{\balpha}{{\mbox{\boldmath$\alpha$}}}
\newcommand{\bnabla}{{\mbox{\boldmath$\nabla$}}}

\newcommand{\be}{\begin{eqnarray}}
\newcommand{\ee}{\end{eqnarray}}
\newcommand{\la}{\langle}
\newcommand{\ra}{\rangle}

\newcommand{\veps}{\varepsilon}

\begin{document}

\title{Nuclear recoil effect on the $g$ factor of highly charged Li-like ions}

\author{V. M. Shabaev}

\affiliation {Department of Physics, St. Petersburg State University,
Universitetskaya 7/9,
199034 St. Petersburg, Russia}

\author{D. A. Glazov}

\affiliation {Department of Physics, St. Petersburg State University,
Universitetskaya 7/9,
199034 St. Petersburg, Russia}

\author{A. V. Malyshev}

\affiliation {Department of Physics, St. Petersburg State University,
Universitetskaya 7/9,
199034 St. Petersburg, Russia}

\author{I. I. Tupitsyn}

\affiliation {Department of Physics, St. Petersburg State University,
Universitetskaya 7/9,
199034 St. Petersburg, Russia}

\begin{abstract}

  The nuclear recoil effect on the $g$ factor of highly charged Li-like ions is evaluated
  in the range $Z=10-92$. The calculations are performed  using the $1/Z$ perturbation theory.
  The one-electron recoil contribution is
  evaluated within the fully relativistic approach  with the wave functions
which account for the screening effects approximately. 
  The two-electron  recoil contributions of the first and higher orders in $1/Z$
  are calculated  within the Breit approximation
  using a four-component approach.

\end{abstract}
\maketitle

\section{Introduction}

Measurements of the $g$ factor of low- and middle-$Z$ H- and Li-like ions
\cite{haf00,ver04,stu11,wag13,lin13,stu13,stu14,koel16}
have reached an accuaracy of a few $10^{-10}$. From the theoretical
side, to get this accuracy we need to evaluate various contributions to the $g$-factor value
\cite{blu97,per97,bei00a,bei00b,sha01,sha02b,nef02,yer02,gla02,sha02c,gla04,lee05,pac05,jen09,vol14,sha15,cza16,yer17a,yer17b,zat17,sha17,mal17,gla18,kar18,cza18}.
The comparison of theory and experiment on the $g$ factors of H- and Li-like
silicon has provided the most stringent tests
of bound-state quantum electrodynamics (QED) in presence of a magnetic field, while
the combination of the experimental and theoretical results on the $g$ factor
of  H-like ions with $Z=6,8,14$ lead to 
the most precise determination of the electron mass \cite{stu14,zat17}.
The measurement of  the isotope shift of the $g$ factor of Li-like
$^{A}$Ca$^{17+}$ with $A=40$ and $A=48$ \cite{koel16} has triggered a special
interest to the calculations of the nuclear recoil contributions to the $g$ factor. 

The fully relativistic theory of the nuclear recoil effect to the first order in
the electron-to-nucleus mass ratio, $m/M$,
on the $g$ factor
of atoms and ions was formulated in Ref. \cite{sha01}, where it was used to derive
closed formulas for the recoil effect on the $g$ factor of H-like ions to all orders in
$\alpha Z$. These formulas remain also valid for an ion with one electron over
closed shells (see, e.g., Ref. \cite{sha02c}), provided the electron propagators 
are redefined for the vacuum with the closed shells included. In that case, in addition
to the one-electron contributions, one obtains two-electron recoil corrections
of the zeroth order in $1/Z$ which can be used to derive  effective four-component
recoil operators within the Breit approximation \cite{sha17}.
The one-electron recoil contribution was evaluated numerically to all
orders in $\alpha Z$ for the $1s$ and $2s$ states in Refs. \cite{sha02b} and \cite{sha17},
respectively. The calculations were performed for the point-nucleus case.
For the ground state of Li-like ions, the two-electron recoil contribution
vanishes to the zeroth order in $1/Z$.  However, the effective recoil
operator can be used to evaluate the recoil corrections of the first
and higher orders in $1/Z$ within the framework of the Breit approximation.
These calculations were carried out for $Z=3-20$ in Ref. \cite{sha17},
where it was found a large discrepancy of the obtained results
with the previous Breit-approximation calculations based on the two-component
approach \cite{yan01,yan02}. As was shown in Ref. \cite{sha17}, this discrepancy
was caused by omitting some important terms in the calculation scheme
formulated within the two-component approach for $s$ states in Ref. \cite{heg75}.
Later  \cite{gla18}, the four-component approach was also used to calculate the recoil effect
within the Breit approximation for middle-$Z$  B-like ions.

Special attention should be paid to probing 
the QED nuclear recoil effect in experiments with heavy ions,
which are anticipated in the nearest future
at the Max-Planck-Institut f\"ur Kernphysik in Heidelberg
and at the HITRAP/FAIR facilities in Darmstadt.
This would provide
an opportunity for tests of QED at strong coupling regime
beyond the Furry picture. To this end, in Ref. \cite{mal17}
the nuclear recoil effect on the $g$ factor of
H- and Li-like Pb and U was calculated and it was shown that
the QED recoil contribution can be probed on a few-percent level
in a specific difference of the $g$ factors of heavy H- and Li-like lead.

In the present paper we extend the calculations of the recoil effect on the $g$ factor of Li-like ions
performed in Refs. \cite{sha17,mal17} to the range $Z=10-92$.  The one-electron
recoil contribution is calculated in the framework of the rigorous QED approach
with the wave functions which partly account for the screening of the Coulomb
potential by the closed shell electrons. As to the two-electron recoil contribution,
it is evaluated within the Breit approximation to all orders in $1/Z$.
All the calculations also partly account for the nuclear size corrections
to the recoil effect.

The relativistic units ($\hbar=c=1$) are used throughout the paper.

%

\section{ Basic formulas}

Let us consider a Li-like ion  
which is put into a 
homogeneous magnetic field, ${\bf A}_{\rm cl}(\bfr)=[
\mbox{\boldmath$\cal H$}
  \times {\bfr}]/2$ with
$ \mbox{\boldmath$\cal H$}$  directed along the $z$ axis.
To zeroth order in $1/Z$, the $m/M$ nuclear recoil contribution to the $g$ factor
is given by a sum of one- and two-electron contributions.  In case of the ground
$(1s)^2 2s$ state the two-electron contribution of zeroth order in $1/Z$
is equal to zero and one has to consider the one-electron term only.
The one-electron recoil contribution to the $g$ factor
 is given by \cite{sha01}
\begin{eqnarray} \label{rec_tot}
\Delta g&=&\frac{1}{\mu_0 m_a}\frac{i}{2\pi M}
\int_{-\infty}^{\infty} d\omega\;
\Biggl[\frac{\partial}{\partial {\cal H}}
\langle \tilde{a}|[p^k-D^k(\omega)+eA_{\rm cl}^k]
\nonumber\\
&&\times\tilde{G}(\omega+\tilde{\veps}_a)
[p^k-D^k(\omega)+eA_{\rm cl}^k]
|\tilde{a}\rangle
\Biggr]_{{\cal H}=0}\,.
\label{06recoilt}
\end{eqnarray}
Here $a$ denotes the one-electron $2s$ state, 
$\mu_0$ is the Bohr magneton, $m_a$ is the angular momentum
projection of the state under consideration, $M$ is the nuclear mass,
 $p^k=-i\nabla^k$ is the momentum operator, 
$D^k(\omega)=-4\pi\alpha Z\alpha^l D^{lk}(\omega)$,
\begin{eqnarray} \label{06photon}
D^{lk}(\omega,{\bf r})&=&-\frac{1}{4\pi}\Bigl\{\frac
{\exp{(i|\omega|r)}}{r}\delta_{lk}\nonumber\\
&&+\nabla^{l}\nabla^{k}
\frac{(\exp{(i|\omega|r)}
-1)}{\omega^{2}r}\Bigr\}\,
\end{eqnarray}
is the transverse part of the photon propagator in the Coulomb 
gauge,  $\balpha$ is a vector
incorporating the Dirac matrices, and
the summation over the repeated indices is implied.  
The tilde sign means that the corresponding quantity
(the wave function, the energy, and the Dirac-Coulomb Green's function
$\tilde{G}(\omega)=\sum_{\tilde{n}}|\tilde{n}\rangle \langle \tilde{n}|[\omega-\tilde{\veps}_n(1-i0)]^{-1}$)
must be calculated in presence of the
magnetic field.

For the practical calculations, the one-electron contribution
is conveniently
represented by a sum of low-order
and  higher-order
terms,
$\Delta g=\Delta g_{\rm L}+\Delta g_{\rm H}$, where 

\be \label{low}
\Delta g_{\rm L}&=&\frac{1}{\mu_0 {\cal H}  m_a}
\frac{1}{M} \la \delta a|
\Bigr[ \bfp^2
 -\frac{\alpha Z}{r}\bigr(\balpha+\frac{(\balpha\cdot\bfr)\bfr}{r^2}\bigr)
  \cdot\bfp
  \Bigr]|a\ra\nonumber\\
&& -\frac{1}{ m_a}\frac{m}{M}\la
a|\Bigl([\bfr \times \bfp]_z -\frac{\alpha Z}{2r}[\bfr \times \balpha ]_z\Bigr)|a\ra\,,
\ee
\be  \label{high}
\Delta g_{\rm H}&=&\frac{1}{\mu_0  {\cal H} m_a}
\frac{i}{2\pi M} \int_{-\infty}^{\infty} d\omega\;\Bigl\{ \la
\delta a|\Bigl(D^k(\omega)-\frac{[p^k,V]}{\omega+i0}\Bigr)\nonumber\\
&&\times G(\omega+\veps_a)\Bigl(D^k(\omega)+\frac{[p^k,V]}{\omega+i0}\Bigr)|a\ra \nonumber\\
&&+\la a|\Bigl(D^k(\omega)-\frac{[p^k,V]}{\omega+i0}\Bigr)
G(\omega+\veps_a)\nonumber\\
&&\times\Bigl(D^k(\omega)+\frac{[p^k,V]}{\omega+i0}\Bigr) |\delta
a\ra\nonumber\\ &&+ \la a|\Bigl(D^k(\omega)-\frac{[p^k,V]}{\omega+i0}\Bigr)
G(\omega+\veps_a)(\delta V-\delta \veps_a)\nonumber\\ &&\times G(\omega+\veps_a)
\Bigl(D^k(\omega)+\frac{[p^k,V]}{\omega+i0}\Bigr)|a\ra \Bigr\}\,. \label{eq3}
\ee
Here
$V(r)$ is the potential of the nucleus or an effective local potential which
is the sum of the nuclear and screening potentials,
 $\delta V(\bfr)=-e\balpha \cdot\bfA_{\rm cl}(\bfr)$,
$G(\omega)=\sum_n|n\ra \la n|[\omega-\veps_n(1-i0)]^{-1}$ is the Dirac-Coulomb Green's
function, $\delta \veps_a=\la a|\delta V|a\ra$, and
$|\delta a\ra=\sum_n^{\veps_n\ne  \veps_a}|n\ra\la n|\delta V|a\ra (\veps_a-\veps_n)^{-1}$.
The low-order term corresponds to the Breit approximation,
while the higher-order term  defines the QED one-electron recoil
contribution.

The recoil contributions of the first and higher orders
in $1/Z$ can be evaluated within the Breit approximation with the use of
the four-component recoil operators \cite{sha17}. The total Breit-approximation
recoil contribution can be represented as a sum of two terms. The first
term is obtained as a combined interaction due to $\delta V$ and 
 the effective recoil Hamiltonian
(see Ref. \cite{sha98} and references therein):
\be \label{br1}
H_M=\frac{1}{2M}\sum_{i,k}\Bigl[\bfp_i\cdot \bfp_k
  -\frac{\alpha Z}{r_i}\bigr(\balpha_i+\frac{(\balpha_i\cdot\bfr_i)\bfr_i}{r_i^2}\bigr)
  \cdot\bfp_k\Bigr]\,.
\ee
The second term is defined by the magnetic recoil 
operator:
\be \label{br2}
H_M^{\rm magn}&=&-\mu_0
\frac{m}{M} \mbox{\boldmath$\cal H$} \cdot
\sum_{i, k}\Bigl\{[\bfr_i\times \bfp_k]\nonumber\\
&&-\frac{\alpha Z}{2r_k}\Bigl[\bfr_i\times\Bigl(\balpha_k
   +\frac{(\balpha_k\cdot\bfr_k)\bfr_k}{r_k^2}\Bigr)\Bigr]
  \Bigr\}\,.
\ee
To zeroth order in $1/Z$, the one-electron parts of the operators (\ref{br1}) and (\ref{br2})
lead to the low-order contribution defined by Eq. (\ref{low}).

Thus, within the four-component Breit-approximation approach
the $m/M$ recoil effect on the $g$ factor can be evaluated
by adding the operators (\ref{br1}) and (\ref{br2})
to the Dirac-Coulomb-Breit Hamiltonian, which includes
the interaction with the external magnetic field.

\section{Numerical calculations}

Let us consider first the calculations within the Breit approxmation.
For these calculations we use  the
operators (\ref{br1}), (\ref{br2}), and the standard
Dirac-Coulomb-Breit Hamiltonian:
\begin{eqnarray} \label{dcb}
H^{\rm DCB}=
\Lambda^{(+)}\Bigl[\sum_{i}h_i^{\rm D}  +\sum_{i<k}V_{ik}\Bigr]\Lambda^{(+)}\,,
\end{eqnarray}
where the indices $i$ and $k$ enumerate the atomic electrons,
$\Lambda^{(+)}$ is the 
projector on the positive-energy states, calculated including
 the interaction with external magnetic field  $\delta V$,
$h_i^{\rm D}$ is the one-electron Dirac Hamiltonian
including $\delta V$, and
\begin{eqnarray} \label{br}
  V_{ik} &=& V^{\rm C}_{ik}+ V^{\rm B}_{ik}\nonumber \\
&=&\frac{\displaystyle \alpha}{\displaystyle r_{ik}} 
-\alpha\Bigl[\frac{\displaystyle{ {\balpha}_i\cdot {\balpha}_k}}
{\displaystyle{ r_{ik}}}+\frac{\displaystyle 1}{\displaystyle 2}
( {\balpha}_i\cdot{\bnabla}_i)({ {\balpha}_k\cdot\bnabla}_k)
r_{ik}\Bigr]
\,
\end{eqnarray}
is the electron-electron interaction operator within the Breit approximation.
The numerical calculations have been performed using the approach based on
the recursive representation of the perturbation theory \cite{gla17}.
The key advantages of the recursive perturbation approach over the standard one
are the universality and the computational efficiency. In Refs. \cite{roz14,var18},
this method was applied to find the higher-order contributions to the Zeeman
and Stark shifts in H-like and B-like atoms. The perfect agreement
of the obtained results with the calculations by other methods was demonstrated.
In Refs. \cite{gla17,mal17a}, the recursive method was used to calculate
the higher-order contributions of the interelectronic interaction in few-electron ions.

The total Breit-approximation recoil contribution for the state under consideration
can be  expressed as
\be \label{ABC}
\Delta g_{\rm Breit}&=&\frac{m}{M}(\alpha Z)^2 \Bigl[A(\alpha Z)+
  \frac{1}{Z} B(\alpha Z)\nonumber\\
&& + \frac{1}{Z^2} C(\alpha Z,Z)\Bigr]\,,
\ee
where the coefficients $A(\alpha Z)$ and $B(\alpha Z)$ 
define the recoil contributions
of the zeroth and first orders in $1/Z$, respectively, while
$C(\alpha Z,Z)$ incorporates the recoil corrections
of the second and higher orders in $1/Z$.
In this work, in the calculation of $C(\alpha Z,Z)$ 
we have taken into account the terms of  the orders $1/Z^2$, $1/Z^3$, and $1/Z^4$. 
The contribution of the terms of  higher orders is much smaller than the present numerical uncertainty.

For the point-nucleus case, 
the  coefficient  $A(\alpha Z)$, which is determined by the one-electron
low-order term (\ref{low}),
can be evaluated
analytically \cite{sha01}. In case of the $2s$ state it is given by
\begin{eqnarray} \label{06shabaeveq112}
  A(\alpha Z)= \frac{1}{4}\frac{8(2\gamma+1)}
  {3(1+\gamma)(2\gamma+\sqrt{2(1+\gamma)}}  \,,
\end{eqnarray}
where $\gamma=\sqrt{1-(\alpha Z)^2}$.
To the leading orders in $\alpha Z$, it leads to
\begin{eqnarray} \label{06shabaeveq113}
A(\alpha Z)=\frac{1}{4} + \frac{11}{192}(\alpha Z)^2+ \cdots \,.
\end{eqnarray}

The calculations to all orders in $1/Z$
have been performed with the point-nucleus recoil operators
defined by Eqs.  (\ref{br1}) and (\ref{br2}) but with the wave functions
evaluated for extended nuclei. This corresponds to a partial treatment 
of the nuclear size corrections to the recoil effect. The Fermi
model of the nuclear charge distribution was used and   
the nuclear charge radii were
taken from Ref. \cite{ang13}. The results of the calculations are presented in
Table \ref{ABC_tab}. The indicated uncertainties are due to the numerical
computation errors.

For the point-nucleus case, the higher-order one-electron contribution (\ref{high})
was calculated for the $1s$ and $2s$ states in a wide range of the  nuclear charge number 
in Refs. \cite{sha02b,sha17}. In the present paper we perform  the calculations
for extended nuclei and effective potentials which partly account for the electron-electron
interaction effects. 
Our first results for  $Z=82,92$
were presented
in Ref. \cite{mal17}, where they were used to search for a possibility
to test QED beyond the Furry picture. In the present paper we extend these calculations to
the range $Z=10-92$.
Since the inclusion of the screening potential into the zeroth-order
Hamiltonian allows one to account for the interelectronic-interaction
effects only partly, we perform the calculations with several
different effective potentials to keep better under control
the uncertainty of the corresponding contribution.
The calculations have been performed for the core-Hartree (CH),
local Dirac-Fock (LDF), and Perdew-Zunger (PZ) effective potentials.
The CH screening potential is derived from the radial charge density of two
$1s$ electrons,
\be
V_{\rm CH}(r)=\alpha\int_0^{\infty} dr'\frac{1}{r_>} \rho_{\rm CH}(r'),
\ee
where $r_>={\rm max}(r,r')$,
\be
 \rho_{\rm CH}(r)=2[G_{1s}^2(r) + F_{1s}^2(r)]\,,\;\;\int_0^{\infty} dr \rho_{\rm CH}(r) =2\,,
\ee
and $G/r$ and $F/r$ are the large and small components  of the radial Dirac wave function.
The LDF potential is constructed
by inversion of the radial Dirac equation with the radial wave functions obtained in the Dirac-Fock
approximation. The corresponding procedure
is described in detail in Ref. \cite{sha05}. 
The last potential
applied in our work is the Perdew-Zunger potential \cite{per81} which
was widely employed in molecular and
cluster calculations.

In Eq. (\ref{high}), the summation over the intermediate electron states
was performed employing 
the finite basis set method.  The basis functions were constructed from B-splines \cite{sap96}
within the framework of the dual-kinetic-balance approach \cite{sha04}.
The integration over $\omega$ 
was carried out analytically for the ``Coulomb'' contribution (the term without
 the $\bfD$ vector) 
and numerically for the ``one-transverse'' and ``two-transverse'' photon contributions
(the terms with one and two $\bfD$ vectors, respectively)
using the Wick's rotation.
The total QED recoil contribution
$\Delta g_{\rm H}$ for the $2s$ state
is conveniently
expressed in terms of the function $P^{(2s)}(\alpha Z)$,
\be
\label{P}
\Delta g_{\rm H}^{(2s)}=\frac{m}{M}\frac{(\alpha Z)^5}{8} P^{(2s)}(\alpha Z)\,.
\ee
The numerical results are presented in  Table~\ref{H-like}.
For comparison, in the second column we list the point-nucleus results which
were partly presented in Ref. \cite{sha17}.

In Table~\ref{F_tab}, we present the total values of the recoil corrections
to the $g$ factor of the ground $(1s)^2 2s$ state of Li-like ions.
They are expressed in terms of the function $F(\alpha Z)$, defined by
\be \label{F}
\Delta g=\frac{m}{M}(\alpha Z)^2 F(\alpha Z)\,.
\ee
The Breit-approximation recoil contributions are obtained by Eq. (\ref{ABC})
with the coefficients given in Table \ref{ABC_tab}.
The uncertainties include both the error bars presented
in Table \ref{ABC_tab} and the uncertainties due to  the
approximate treatment  of the nuclear size correction to
the recoil effect. We have assumed that the relative value of the latter uncertainty is equal to the related
contribution to the binding energy which was evaluated within the Breit approximation in Ref. \cite{ale15}.
For the QED recoil contribution we use
the LDF values from Table~\ref{H-like}.
The  uncertainty of this term
is estimated as a sum of two contributions.
The first one is due to  the approximate treatment of the electron-electron
interaction effect on the QED recoil contribution.
This uncertainty was estimated by performing the calculations of the low-order
(non-QED) one-electron recoil contribution
with the LDF potential and comparing the obtained result with the total Breit recoil value
evaluated above. 
The ratio of the obtained difference  to the non-QED LDF result was chosen 
as the relative uncertainty of the corresponding correction to the
QED recoil contribution. It should be noted that this uncertainty exceeds
 the difference between the results obtained for the different
screening potentials presented in Table~\ref{H-like}.
The second contribution to the uncertainty is 
caused by the approximate treatment of the nuclear size correction to the recoil effect.
It was estimated in the same way as for the Breit recoil contribution.
As one can  see from Table~\ref{F_tab}, for very heavy ions the QED recoil effect
becomes even bigger than the Breit recoil contribution.

The total recoil contribution to the $g$ factor
should also include small corrections of orders $\alpha (\alpha Z)^2(m/M)$
and $ (\alpha Z)^2(m/M)^2$ and the related corrections of higher
orders in $\alpha Z$ and in $1/Z$.
To the lowest order in $\alpha Z$ the corresponding one-electron corrections were evaluated 
in Refs. \cite{gro71,clo71,pac08,eid10}.

\section{Conclusion}

In this paper we have evaluated the nuclear recoil effect of the first
order in $m/M$  on the ground-state $g$ factor of highly charged Li-like ions.
The Breit-approximation contributions have been calculated to all orders
in $1/Z$ employing the recursive perturbation theory. The one-electron
higher-order (QED) recoil contribution was evaluated to all orders in $\alpha Z$ 
with the wave functions which partly account for the electron-electron
interaction effects. As the result, the most precise theoretical
predictions for the recoil effect on the  $g$ factor of highly charged
Li-like ions are presented.


\section{Acknowledgments}
This work was supported by the Russian Science Foundation (Grant No. 17-12-01097).
%
%

\begin{table}
  \caption{The Breit-approximation recoil contributions to the $g$ factor of the
    $(1s)^2 2s$ state of  Li-like ions 
    expressed in terms of
    the coefficients    $A(\alpha Z)$,  $B(\alpha Z)$, and  $C(\alpha Z, Z)$  defined
    by  Eq. (\ref{ABC}).}
\label{ABC_tab}
\begin{tabular}{cr@{.}lr@{.}lr@{.}lr@{.}lr@{.}l} \hline
$Z$& \multicolumn{2}{c}{$A(\alpha Z)$}
                    & \multicolumn{2}{c}{$B(\alpha Z)$}
  & \multicolumn{2}{c}{$C(\alpha Z, Z)$ }
\\
\hline
10  &   0&2503  &   $-$0&5172    &  $-$0&236(4)  \\
12  &   0&2504  &   $-$0&5179    &  $-$0&243(3)  \\
14  &   0&2506  &   $-$0&5187    &  $-$0&245(3)  \\
16  &   0&2508  &   $-$0&5197    &  $-$0&248(3)  \\
18  &   0&2510  &   $-$0&5207    &  $-$0&250(2)  \\
20  &   0&2512  &   $-$0&5219    &  $-$0&250(2)  \\
24  &   0&2517  &   $-$0&5247    &  $-$0&250(1)  \\
28  &   0&2524  &   $-$0&5279    &  $-$0&247(1)  \\
30  &   0&2527  &   $-$0&5297    &  $-$0&245(1)  \\
32  &   0&2531  &   $-$0&5315    &  $-$0&243  \\
40  &   0&2548  &   $-$0&5402    &  $-$0&228  \\
48  &   0&2567  &   $-$0&5504    &  $-$0&205  \\
50  &   0&2572  &   $-$0&5531    &  $-$0&198  \\
56  &   0&2588  &   $-$0&5618    &  $-$0&177  \\
60  &   0&2599  &   $-$0&5677    &  $-$0&160  \\
64  &   0&2607  &   $-$0&5734    &  $-$0&141  \\
70  &   0&2616  &   $-$0&5813    &  $-$0&105  \\
72  &   0&2617  &   $-$0&5836    &  $-$0&092  \\
80  &   0&2606  &   $-$0&5886    &  $-$0&037  \\
82  &   0&2597  &   $-$0&5881    &  $-$0&019  \\
90  &   0&2510  &   $-$0&5721    &  0&051  \\
92  &   0&2471  &   $-$0&5629    &  0&065  \\

\hline
\end{tabular}
\end{table}

\begin{table}
\caption{The higher-order (QED) recoil contribution to the $2s$ $g$ factor
expressed in terms of the function $P^{(2s)}(\alpha Z)$ defined by
Eq. (\ref{P}).
The indices Coul, CH, LDF and PZ refer to the Coulomb and various screening potentials, see the text.
The indices p.n. and f.n. correspond to the point-like and finite-size nuclear models.}
\label{H-like}
\begin{tabular}{cr@{.}lr@{.}lr@{.}lr@{.}lr@{.}l} \hline
  $Z$& \multicolumn{2}{c}{$P^{\rm (p.n.)}_{\rm Coul}(\alpha Z)$}
& \multicolumn{2}{c}{$P^{\rm (f.n.)}_{\rm Coul}(\alpha Z)$}
                    & \multicolumn{2}{c}{$P^{}_{\rm CH}(\alpha Z)$}
                                     & \multicolumn{2}{c}{$P^{}_{\rm LDF} (\alpha Z)$ }
                                                     & \multicolumn{2}{c}{$P^{}_{\rm PZ}(\alpha Z)$}
\\
\hline
10 & 8&8762(1)  &   8&9145   &   6&2670      &   6&1840   &    6&6098  \\

12 & 8&1943(1)  &   8&2333   &   6&1987      &   6&1326   &    6&4787  \\

14 & 7&6447(1)  &   7&6847   &   6&0614      &   6&0069   &    6&2955  \\

16 & 7&1911(1)  &   7&2325   &   5&8998      &   5&8539   &    6&0995  \\

18 & 6&8101(1)  &   6&8539   &   5&7349      &   5&6953   &    5&9081  \\

20 & 6&4860(1)   &   6&5309   &   5&5740   &    5&5395     &    5&7264   \\

24 & 5&9670(1)   &   6&0151   &   5&2844   &    5&2571     &    5&4065   \\

28 & 5&5753(1)   &   5&6267   &   5&0429   &    5&0205     &    5&1446   \\

30 & 5&4160(1)   &   5&4703   &   4&9412   &    4&9208     &    5&0351   \\

32 & 5&2771(1)   &   5&3341   &   4&8509   &    4&8322     &    4&9382   \\

40 & 4&8840(1)   &   4&9487   &   4&5900  &    4&5760     &    4&6588   \\

48 & 4&6937(1)   &   4&7686   &   4&4789  &    4&4680     &    4&5375   \\

50 & 4&6727(1)    &   4&7499   &   4&4723  &    4&4619     &    4&5290   \\

56 & 4&6697(1)    &   4&7530   &   4&5028  &    4&4940     &    4&5557   \\

60 & 4&7182(1)  &   4&8035   &   4&5658  &    4&5578     &    4&6171   \\

64 & 4&8098(2)  &   4&8958   &   4&6670  &    4&6596     &    4&7173   \\

70 & 5&039(1)  &   5&1144   &   4&8928  &    4&8865     &    4&9429   \\

72 & 5&144(1)  &   5&2114   &   4&9908  &    4&9847     &    5&0411  \\
 
80 & 5&753(3)  &   5&7437   &   5&5200  &    5&5152     &    5&5728   \\

82 & 5&965(3)  &   5&9188   &   5&6926  &    5&6881     &    5&7464   \\

90 & 7&19(2)  &   6&8284   &   6&5850  &    6&5818     &    6&6449   \\

92 & 7&64(2)  &   7&1187   &   6&8689  &    6&8662     &    6&9309   \\

\hline

\end{tabular}
\end{table}

\begin{table}
  \caption{The Breit, QED, and total recoil contributions to the $g$ factor of the
    $(1s)^2 2s$ state of  Li-like ions 
    expressed in terms of
    the function $F(\alpha Z)$ defined by Eq. (\ref{F}).
  }
\label{F_tab}
\begin{tabular}{cr@{.}lr@{.}lr@{.}lr@{.}lr@{.}l} \hline
$Z$& \multicolumn{2}{c}{$F_{\rm Breit}$}
                    & \multicolumn{2}{c}{$F_{\rm QED}$}
  & \multicolumn{2}{c}{$F_{\rm total}$ }
\\
\hline
10  &   0&1962(1)  &   0&0003(1)    &  0&1965(1)  \\
12  &   0&2056  &   0&0005(1)    &  0&2061(1)  \\
14  &   0&2123  &   0&0008(1)    &  0&2131(1)  \\
16  &   0&2173  &   0&0012(1)    &  0&2185(1)  \\
18  &   0&2213  &   0&0016(2)    &  0&2229(2)  \\
20  &   0&2245  &   0&0022(2)    &  0&2266(2)  \\
24  &   0&2294  &   0&0035(3)    &  0&2330(3)  \\
28  &   0&2332  &   0&0054(3)    &  0&2385(3)  \\
30  &   0&2348  &   0&0065(4)    &  0&2412(4)  \\
32  &   0&2362  &   0&0077(4)    &  0&2439(4)  \\
40  &   0&2411  &   0&0142(6)    &  0&2553(6)  \\
48  &   0&2452  &   0&0240(8)    &  0&2692(8)  \\
50  &   0&2461  &   0&0271(9)    &  0&2732(9)  \\
56  &   0&2487(1)  &   0&0383(11)    &  0&2871(11)  \\
60  &   0&2503(1)  &   0&0478(13)    &  0&2982(13)  \\
64  &   0&2517(2)  &   0&0593(15)    &  0&3110(16)  \\
70  &   0&2533(3)  &   0&0814(20)    &  0&3347(20)  \\
72  &   0&2536(4)  &   0&0904(22)    &  0&3440(22)  \\
80  &   0&2533(10)  &   0&1372(30)    &  0&3904(32)  \\
82  &   0&2525(12)  &   0&1523(33)    &  0&4048(36)  \\
90  &   0&2446(28)  &   0&2331(54)    &  0&4777(61)  \\
92  &   0&2410(35)  &   0&2597(65)    &  0&5007(73)  \\

\hline
\end{tabular}
\end{table}



\end{document}